# Accelerometer Based Method for Tire Load and Slip Angle Estimation

Kanwar Bharat Singh and Saied Taheri
Center for Tire Research (CenTiRe)
Department of Mechanical Engineering, Virginia Tech, Blacksburg VA 24061, USA

**Abstract:** Tire mounted sensors are emerging as a promising technology, capable of providing information about important tire states. This paper presents a survey of the state-of-the-art in the field of smart tire technology, with a special focus on the different signal processing techniques proposed by researchers to estimate the tire load and slip angle using tire mounted accelerometers. Next, details about the research activities undertaken as part of this study to develop a smart tire are presented. Finally, novel algorithms for estimating the tire load and slip angle are presented. Experimental results demonstrate the effectiveness of the proposed algorithms.

**Keywords:** smart tire system, signal processing, tire load, tire slip angle, accelerometer

## 1. Introduction

The application of sensor technology in tires has been extensively studied over the last decade by many researchers [1-6]. The stimulus for the research and application of sensor technology in tires was the Bridgestone/Firestone recall of 14.4 million tires in 2000 [7]. Because of the recalls, U.S. legislators included tire pressure monitoring (TPMS) as part of its TREAD Act [8-10]. Thereafter, European countries (via United Nations regulation UNECE-R64) adopted the TPMS legislation for new models beginning in November 2012, and for all vehicles beginning in November 2014. South Korea has adopted similar TPMS legislation to that of Europe. Likewise, countries such as Japan, China and India are in the process of investigating TPMS technology as well.

Although classical TPMS sensors are mounted on the valve stem inside the wheel, there is a great level of interest in placing the sensor directly on the tire. A tire mounted sensor will not only fulfill the basis TPMS functionalities, i.e. measure cavity air pressure and temperature, but also presents opportunities to sense key attributes related to the tire contact patch. This would be achieved through the inclusion of an additional sensing element (e.g. an accelerometer) and advanced digital signal processing (DSP) algorithms. Running sophisticated DSP algorithms requires advanced low power microchips. The internet of things (IoT) digital revolution is resulting in an exponential growth in computing power of embedded microchips thus giving great impetus to the development of Smart Tire Technology. Large scale usage of IoT sensors across several industries is driving the economies of scale. The availability of a low-cost sensing solutions becomes more critical in the case of tires, knowing they are largely a commoditized product.

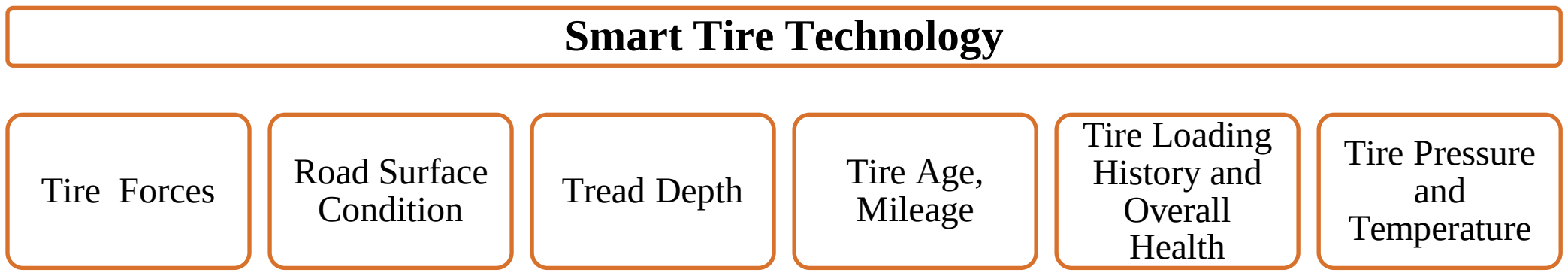

**Figure 1: Smart tire system sensor output**

Throughout this paper, the term Smart Tire Technology explicitly refers to a configuration wherein the sensor is mounted on the tire inner liner.

The Smart Tire is expected to have capabilities to monitor in real-time the tire forces, road surface conditions, tire slip conditions, temperature, pressure and the tire tread depth (Figure 1). Being able to leverage a plethora of information, a Smart Tire is expected to stimulate the development of advanced traction, braking and stability control systems for improving vehicle safety and performance [11-14].

There have been several architectures proposed in the past few years pertaining to Smart Tires (Figure 2):

(i) The APOLLO program - Developed a 3-in-l intelligent tire [15];
(ii) The FRICTI@N program - On-board system for measuring and estimating tire-road friction [16];
(iii) Contact area information sensing (CAIS) system being developed by Bridgestone [17];
(iv) Cyber Tire being developed by Pirelli Tires [18];
(v) Intelligent tire system being developed by Continental AG [19];
(vi) Intelligent tire system being developed by Hyundai Motor Co, Mando Corp, and Corechips [20].
(vii) IntelliGrip Concept from The Goodyear Tire & Rubber Company [21]

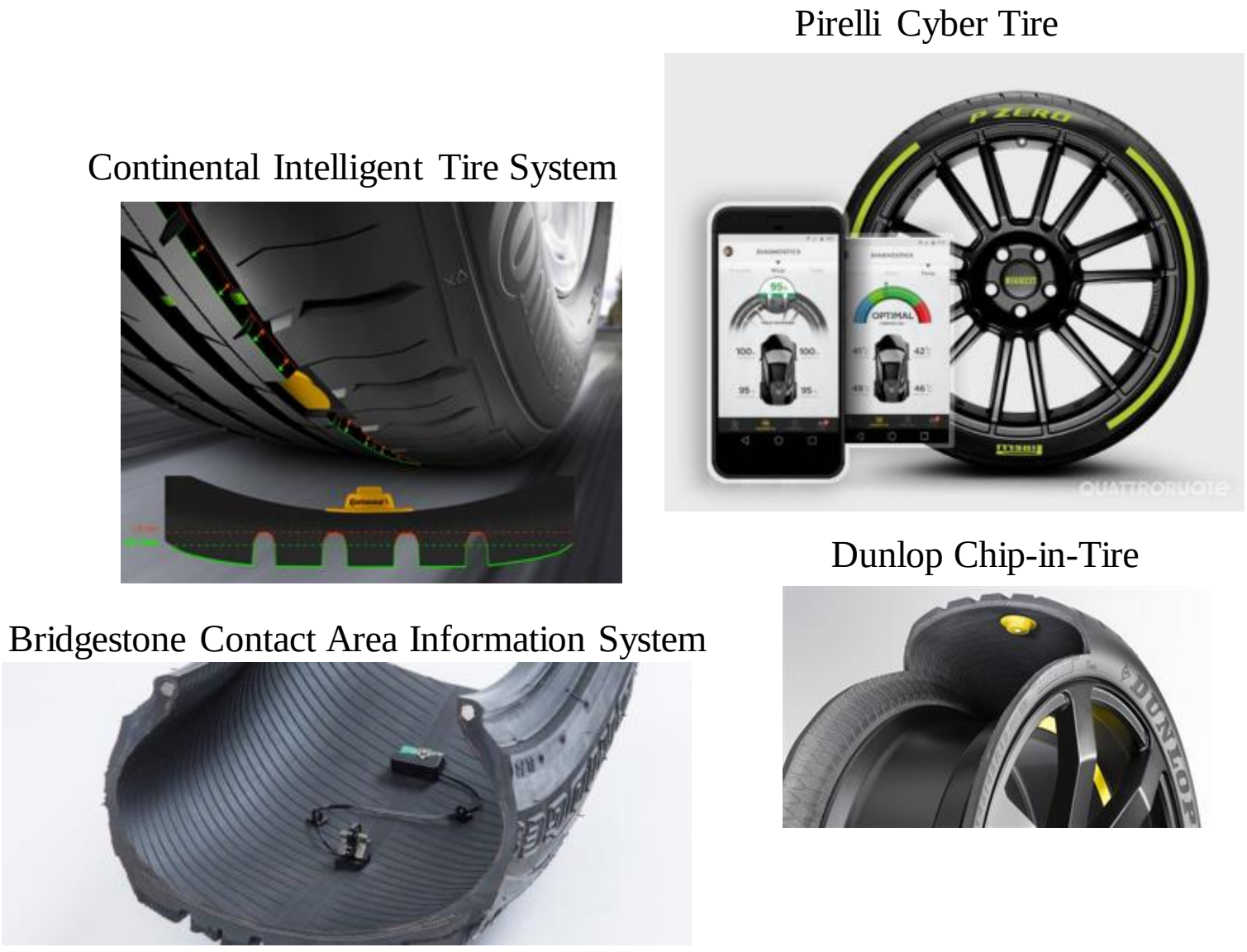

**Figure 2: Smart tire system – products and concepts**

The key value proposition for usage of smart tires in vehicles is the potential they offer in improving performance of vehicle control systems. This would be of even more relevance in the case of autonomous vehicles (AVs), considering their high safety requirements.

This paper first presents a comprehensive overview of the state-of-the-art in the field of Smart Tire Technology and thereafter presents details of the work undertaken at the Center for Tire Research (CenTiRe) to develop a smart tire.

The main contributions of this paper are as follows:

- Literature review of the different signal processing techniques used by researchers to extract meaningful information from tire mounted sensors, specifically from tire mounted accelerometers.
- Details about research activities undertaken as part of this study to develop a smart tire.
- Novel algorithms for estimating the tire load and slip angle.

## 2. Literature Review

There are several signal processing algorithms proposed in literature for estimating tire load and slip angle. Table 1 presents a summary of the different techniques proposed by researchers.

**Table 1: State-of-the-art literature review – analysis of tire accelerometer data**

| Estimated state | Signal used | Underlying physics | Reference |
|---|---|---|---|
| Tire vertical force | Radial acceleration | Use an empirical model between tire contact patch length and tire load | US Patent 7404317 [22]<br>US Patent 7536903 [23]<br>US Patent 8742911 [24]<br>US Patent 20130261991 [25] |
| | Radial acceleration | Use an empirical model correlating the radial acceleration amplitude to the tire load | US Patent 8874386 [26] |
| | Radial acceleration | Use an empirical model to describe the shape of the radial acceleration signal - Vertical load is treated as an unknown parameter and is estimated used an EKF observer | Teerhuis et al. 2015 [27] |
| | Radial acceleration | Use Principal Component Analysis, to describe smart tire signal variance by means of a few linear projections. Projections are correlated linearly with the contact forces | Krier et al. 2014 [28] |
| Tire slip angle | Radial acceleration | Monitor change in the footprint length at the centerline and shoulder positions to estimate the tire slip angle | US Patent 7552628 [29] |
| | Lateral acceleration | Correlate lateral acceleration amplitude with the tire slip angle | US Patent 8024087 [30] |
| | Lateral acceleration | Extract lateral deformation from the acceleration signal and correlate that with the tire slip angle | Matsuzaki et al. [31] |

In view of the above literature review, the following conclusions can be drawn:

- Among the three axes, the radial acceleration signal has been exploited the most by researchers.
- Most of these applications require data sampled at high rates. For instance, for tire-road friction sensing, sensor signal needs to be sampled at 5-10 kHz.
- Both time and frequency domain signal processing techniques have been extensively used.
- Current tire pressure monitoring (TPMS) chips available in production vehicles cannot support such complex signal processing algorithms. Hence, an application specific IC would to be required.
- Since these sensor systems are expected to survive the tire service life, they would require a power source capable of supporting high sampling rate data processing and high frequency data transmission.

## 3. Development of a Smart Tire System

Based on their proven reliability for tire application, it was decided to use tri-axial micro electro mechanical systems (MEMS) accelerometers in this study (Table 2). These sensors offer the advantage of their small dimension, low weight, high reliability, robustness and reliable performance even in harsh and hostile environments.

**Table 2: Tri-axial accelerometer characteristics**

| 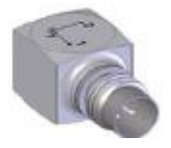 | **Accelerometer Characteristics** |
|---|---|
| Range | ± 5000 g |
| Sensitivity | 1 mv/g |
| Frequency response | 1.2 -10 kHz |
| Resonance frequency | 30 kHz |
| Mass | 3.0 g |
| Dimensions | 17.7 x 9.02 x 9.14 mm |

The smart tire was developed by placing the tri-axial accelerometer on the inner liner of a tire (Figure 3a). Figure 3b shows the final assembly of the instrumented tire with a high-speed slip ring attached to the wheel. Extensive on-road tests were conducted using the in-house mobile tire test rig shown in Figure 3c.

(a) (c)

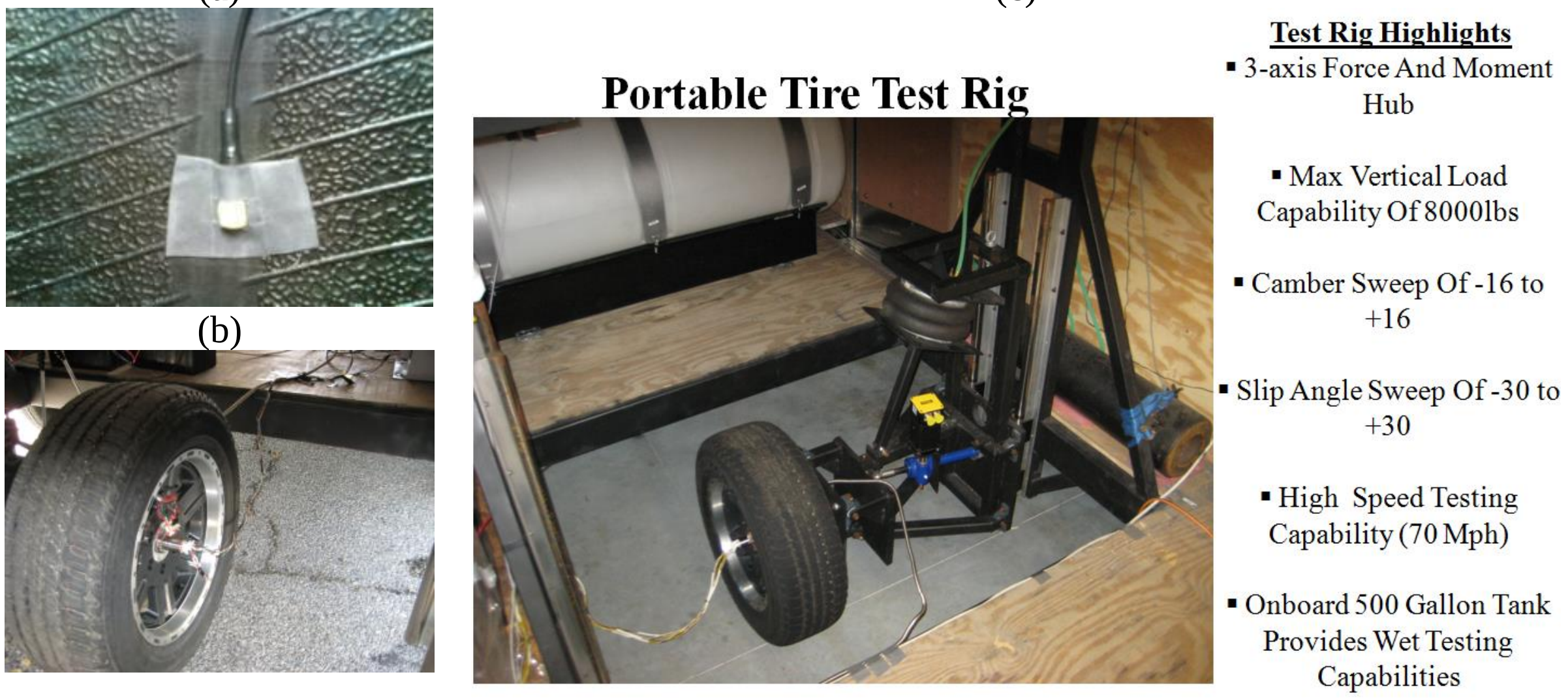

**Figure 3: Smart tire application: (a) sensor mounting location, (b) instrumented tire assembly, (c) mobile tire test rig**

An example of the measured acceleration signal for one wheel turn is shown in Figure 4.

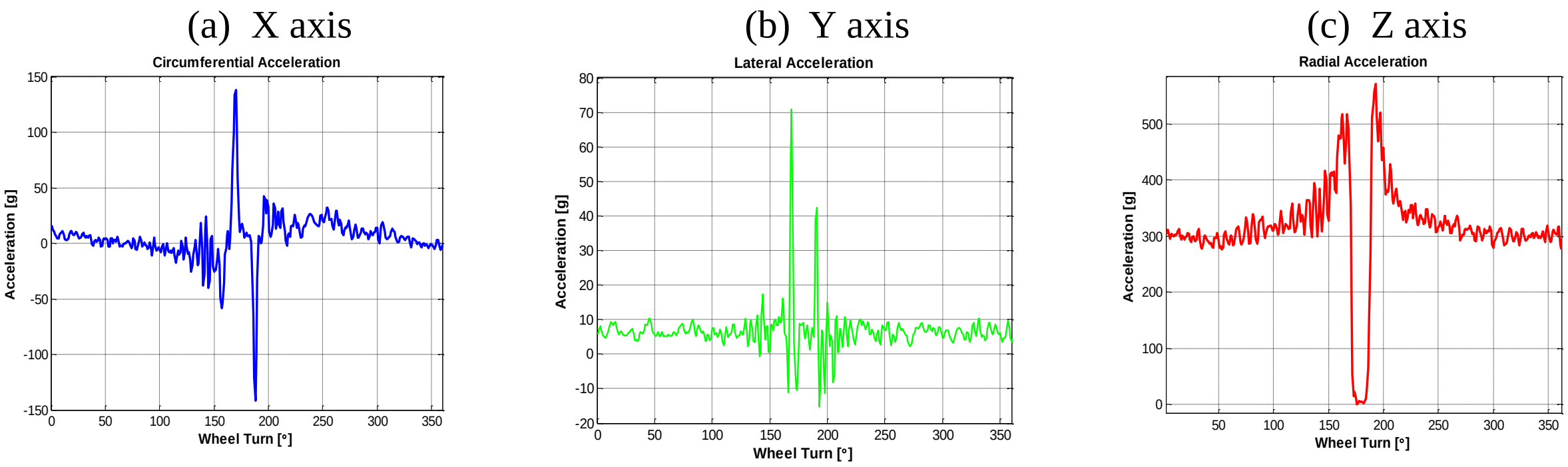

**Figure 4: Signal for one wheel turn from a tire attached tri-axial accelerometer.**

Tire load and slip angle estimation were treated as the lead applications for this study. The motivation for considering these tires states is twofold:

- First, with information about individual tire loads from tire mounted sensors, the vehicle inertial parameters, namely, vehicle mass, yaw moment of inertia and axle distances from the vehicle center of gravity, can be precisely estimated, thus making the vehicle controller robust against variations in vehicle model parameters.
- Second, it is well known in literature that vehicle sideslip angle can significantly reduce the risk of accidents through effective design and implementation of advanced chassis control systems. However, in production vehicles, sideslip angle is difficult to measure within the desired accuracy level because of high costs and other associated impracticalities. Having a direct measurement of the tire slip angle from a smart tire would enable the design of novel control systems using tire slip angle as a feedback signal instead of vehicle sideslip angle.

Specific details of the signal processing algorithms developed as part of this study are given in the subsequent subsections.

## 4. Algorithm Details

### 4.1 Tire load estimation

A common approach proposed in literature to estimate the tire load is to exploit the relationship between load and the tire contact patch length. A parametric study was conducted to quantify the influence of load and other relevant parameters on the tire contact patch length. The patch length was extracted from the tangential acceleration signal by identifying the leading and trailing edge of the tire using a peak detection algorithm.

As shown in Figure 5, tire load and inflation pressure were seen to have a significant impact on the patch length of the tire.

- Higher load results in a longer contact patch length
- Higher inflation pressure results in shorter contact patch length

These results are in-line with results previously published in literature.

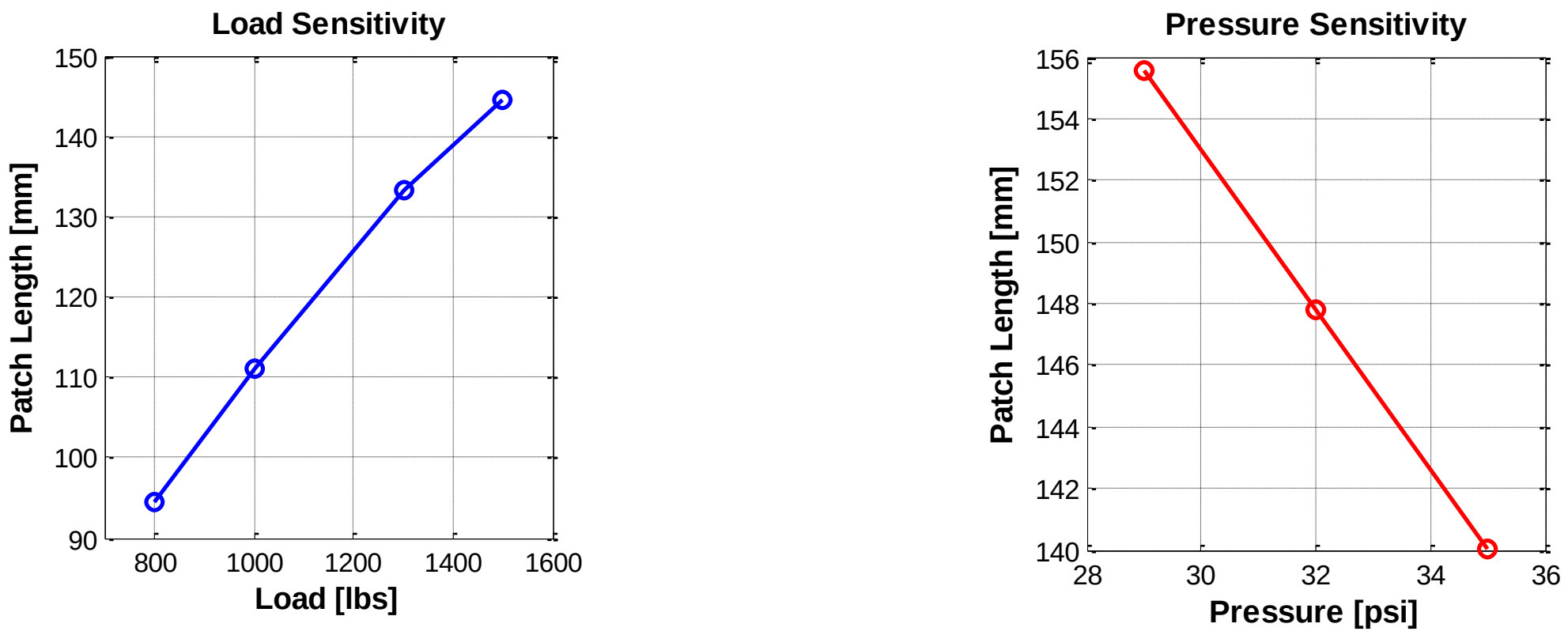

**Figure 5: Impact of load and inflation on the tire contact patch length**

What was not so unexpected and in fact not captured in published literature is the strong correlation seen between the tire wear state (i.e. remaining tread depth) and the tire contact patch length (Figure 6). Lowering the tread depth results in a decrease in the tire circumference (i.e. a shrinkage of the tire radius) and consequently a decrease in the contact patch length.

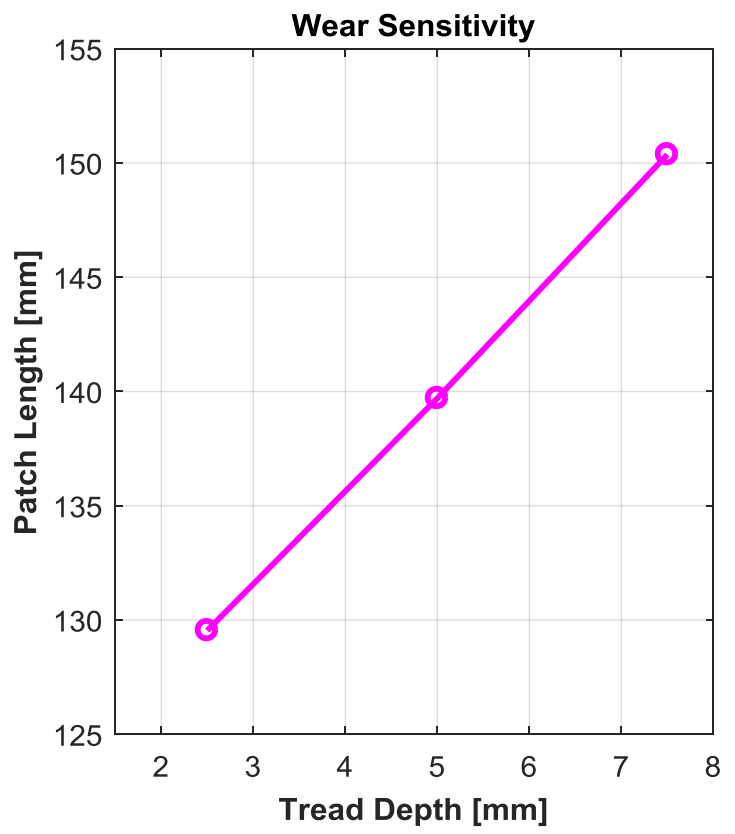

**Figure 6: Impact of remaining tread depth on the tire contact patch length**

This poses an issue for the load estimation model. The load model will need to have knowledge of the tire wear state, which by itself presents several technical challenges considering the complexity of typical models required for estimating the remaining tread depth.

To overcome this challenge, alternative signal features showing a strong correlation with the tire loading state were studied. Knowing that a tire deflects vertically in a linear manner as the load is increased, it was decided to extract the peak radial displacement of the tire from the radial acceleration signal. This requires double integrating the acceleration signal to retrieve displacement. Unfortunately, accelerometers have an unwanted phenomenon called drift, caused by a small DC bias in the acceleration signal. The presence of drift can lead to large integration errors. If the acceleration signal from a real accelerometer is integrated without any filtering performed, the output could become unbounded over time. To solve the problem of drift, a high-pass filter was used to remove the DC component of the acceleration signal. This is done by carefully extracting the acceleration signal per wheel turn and then applying the high pass filter in conjunction with the integration operation (Figure 7).

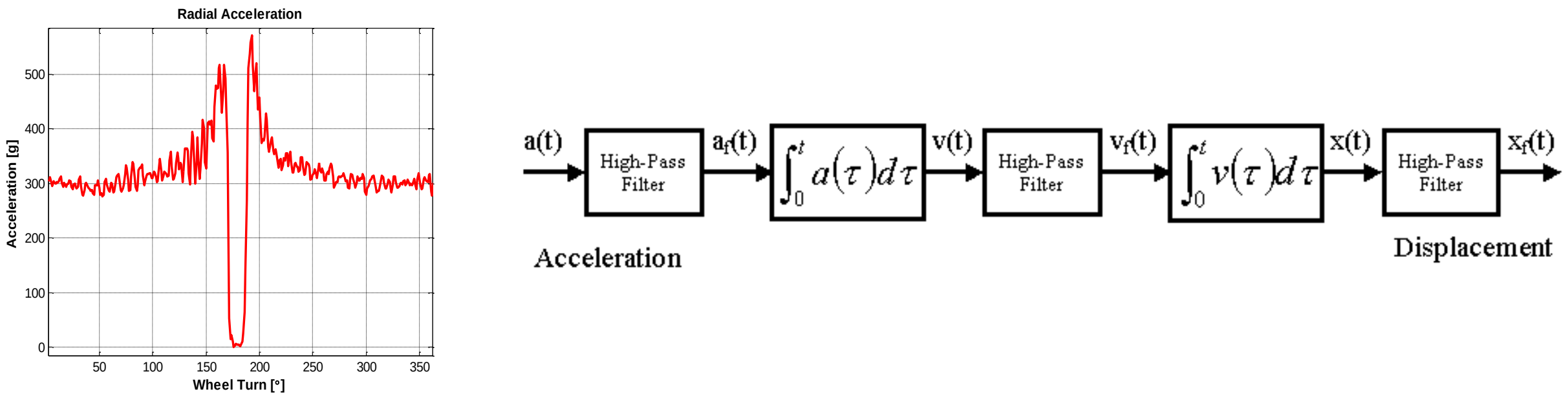

**Figure 7: Algorithm for retrieving radial displacement from the acceleration signal**

By filtering before integrating, drift error was eliminated, and the radial displacement successfully extracted (Figure 8).

(a) (b)

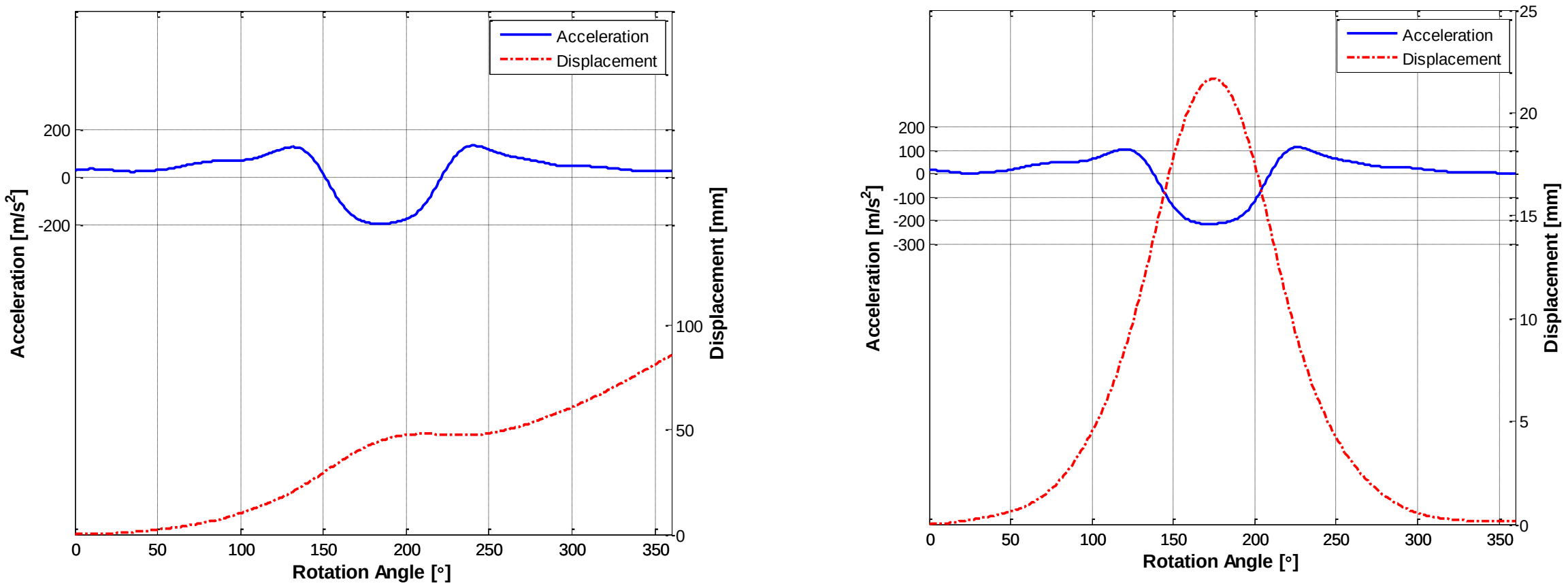

**Figure 8: (a) Integration without filtering, (b) Integration with filtering**

Finally, the amplitude of the peak radial displacement of the tire was extracted from the displacement curve. The radial displacement shows near linear correlation with the tire load and inflation pressure (Figure 9).

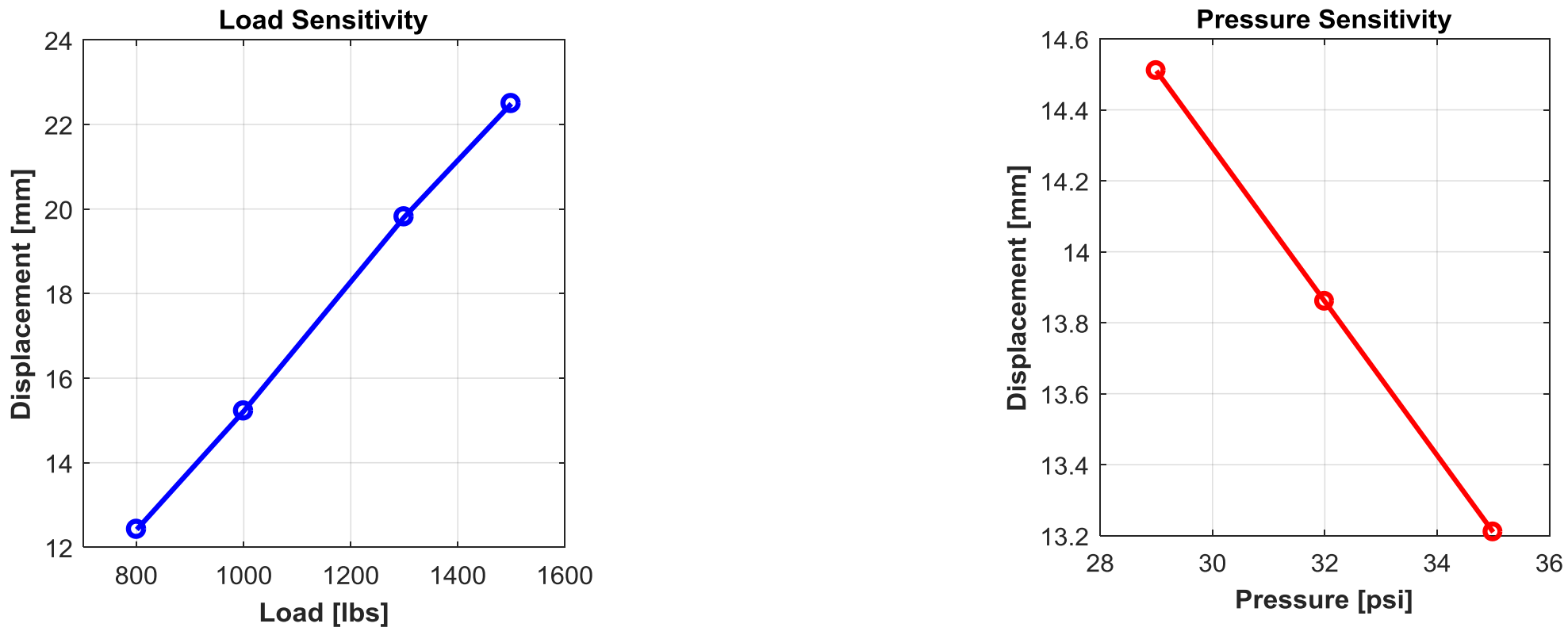

**Figure 9: Impact of load and inflation on the tire vertical displacement**

More interestingly, in comparison to the tire contact patch length, amplitude of the peak radial displacement showed a much higher sensitivity to tire load (Table 3). Moreover, radial displacement shows negligible sensitivity to the tire tread depth. This makes radial displacement a more attractive feature in comparison to contact patch length and hence was the preferred feature for this study.

Table 3: Sensitivity summary

| | **Load** | **Pressure** | **Tread Depth** |
|---|---|---|---|
| Range Tested | [800-1500 lbs.] | [29-35 psi] | [2mm -8mm] |
| Peak Radial Displacement | 80-90% | 10-15% | negligible |
| Contact Patch Length | 40-50% | 10-15% | 15-20% |

A model capturing the dependencies between the tire radial displacement, load and inflation pressure was fit (Figure 10). A polynomial fit with second-order in pressure and first-order in load provided a good model fit.

Peak Displacement= p00 + p10*load + p01*pressure + p11*load*pressure + p02*pressure^2 (1)

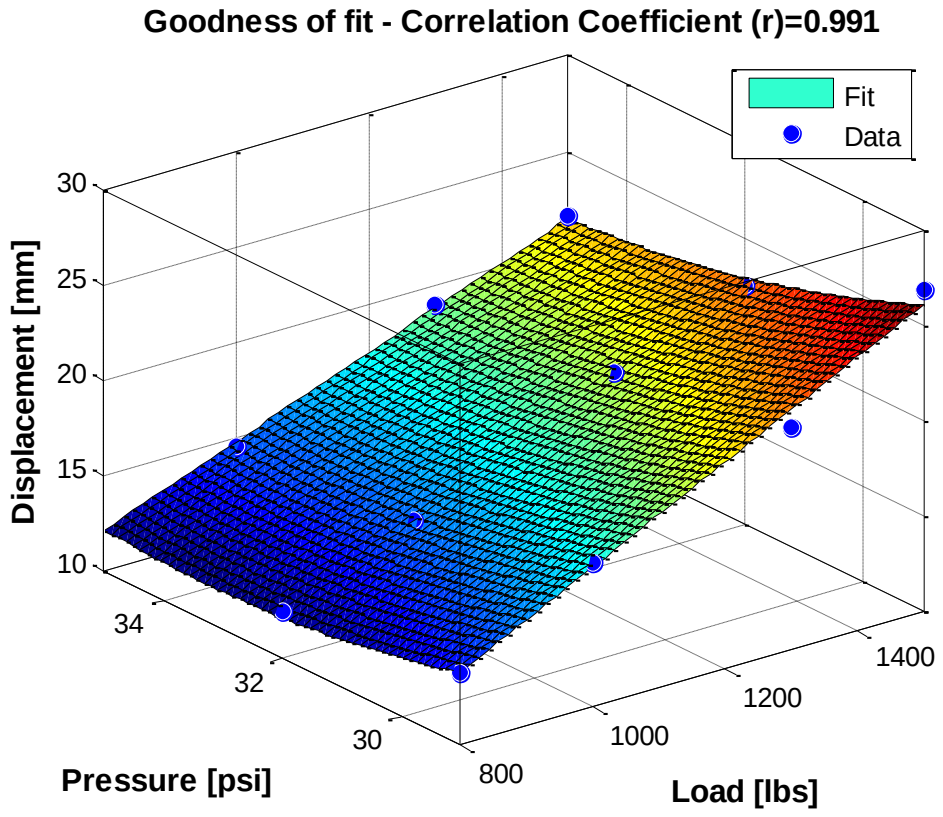

**Figure 10: Model fitting**

Equation (1) can be rewritten into a standard parameter identification form as follows:

$$y(t) = \varphi^T(t) \cdot \theta(t)$$

$$y(t) = \textit{Peak Displacement- p00 - p10*pressure - p20*pressure\^{}2)/ (p11*pressure + p01)} \quad (2)$$
$$\varphi^T(t) = 1$$
$$\theta(t) = \textit{unknown variable (tire load)}$$

The unknown parameter $\theta(t)$ can be identified in real-time using parameter identification approach. The recursive least squares (RLS) algorithm [52] provides a method to iteratively update the unknown parameter at each sampling time to minimize the sum of the squares of the modeling error using the past data contained within the regression vector, $\varphi(t)$. The performance of the RLS algorithm was evaluated through experimental road vehicle tests. The vehicle maneuver was straight driving with intermittent gas pedal presses.

(a) Raw Measurement – Straight-line driving

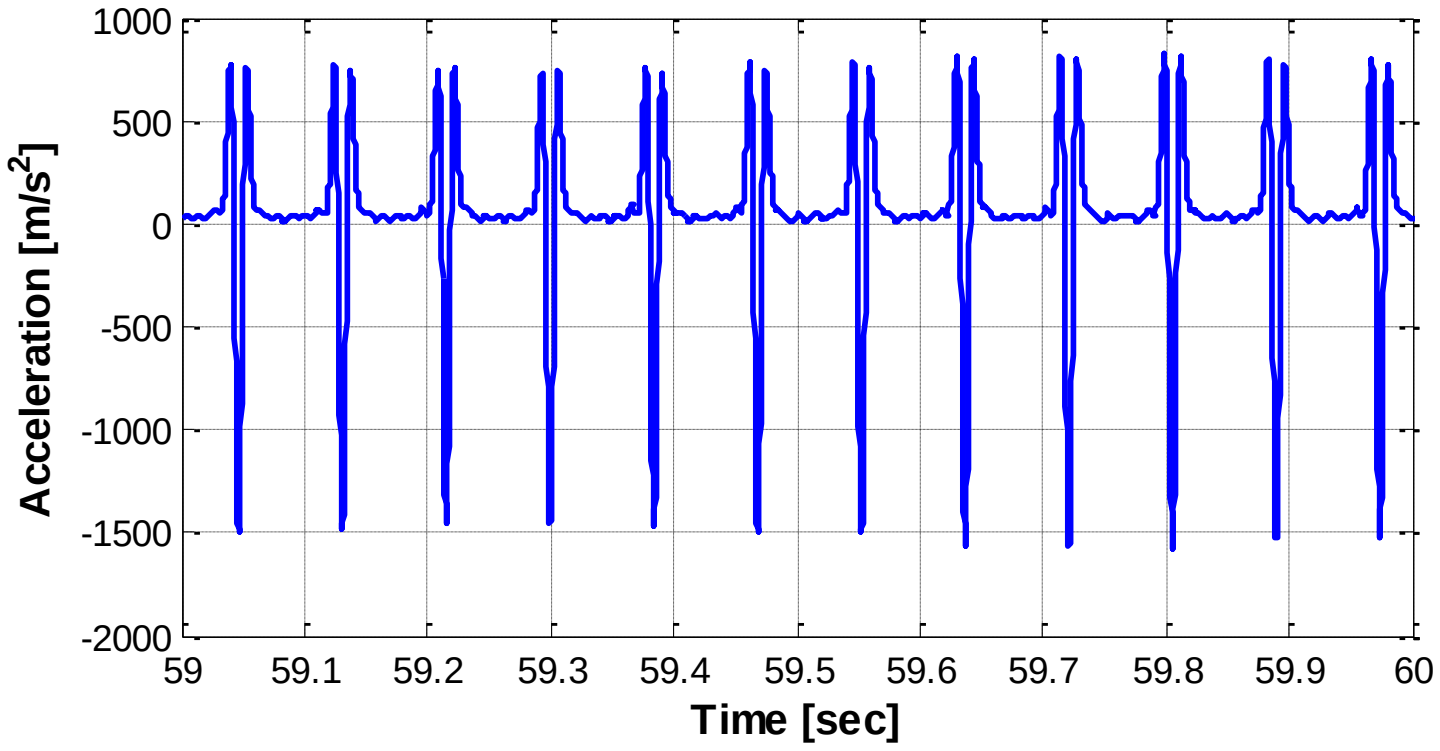

(b) Feature Extraction

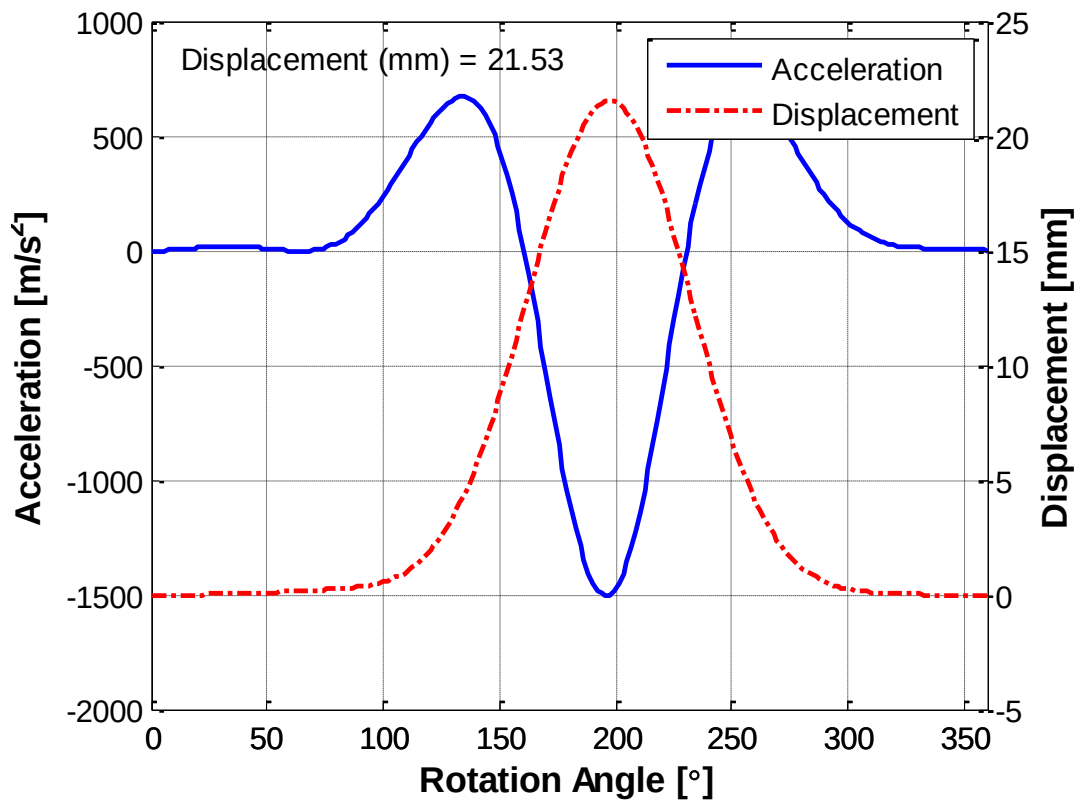

(c) RLS Prediction model

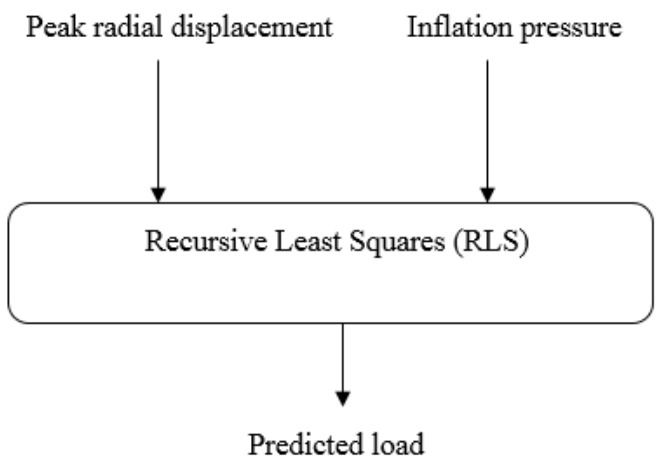

Predicted tire load 1490 lbs.
Actual Tire Load (measured)= 1455 lbs.
Error=2.6%

**Figure 11: Prediction Performance**

The RLS model was found to predict the tire load within the first 20 wheel rotations. The estimation error was 2.6%. For the sake of completeness, the same dataset was also used to evaluate the performance of a contact patch length based load estimation model. The estimation error was found to be 5.3%. So, a radial deflection based model was found to superior than the contact patch length based model.

## 4.2 Tire slip angle estimation

Using the procedure described in Figure 7 , the lateral displacement profile of the tire footprint was extracted from the lateral acceleration signal at different tire slip angles (Figure 12).

The following observations were made:

- the maximum lateral displacement of the tire footprint increases as the slip angle increases.
- the slope of initial linear part of the lateral displacement profile also increases as the slip angle increases.

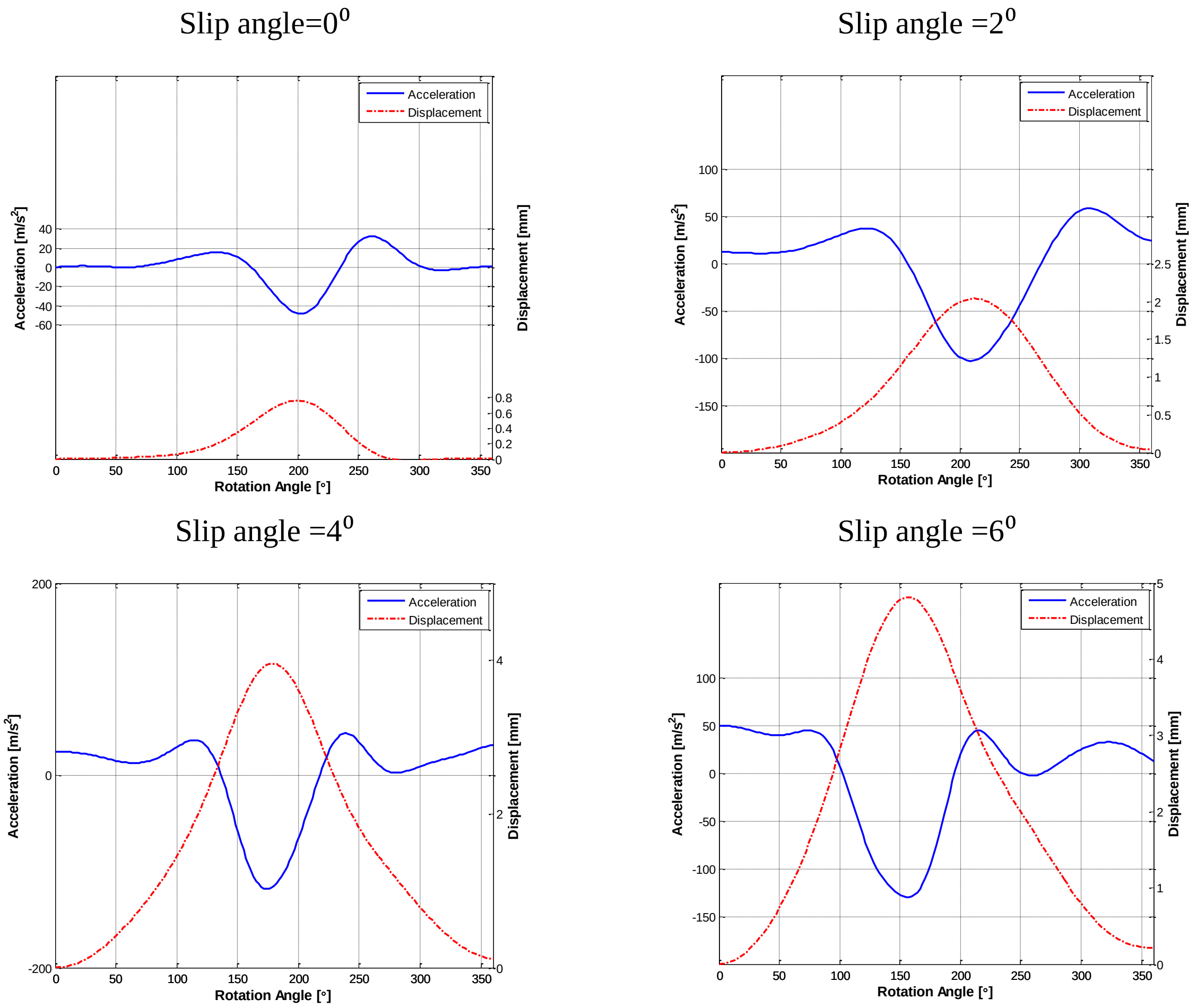

**Figure 12: Extraction of lateral displacement from the lateral acceleration signal at different tire slip angles**

Both these features were founded to be linearly correlated with the tire slip angle (Figure 13).

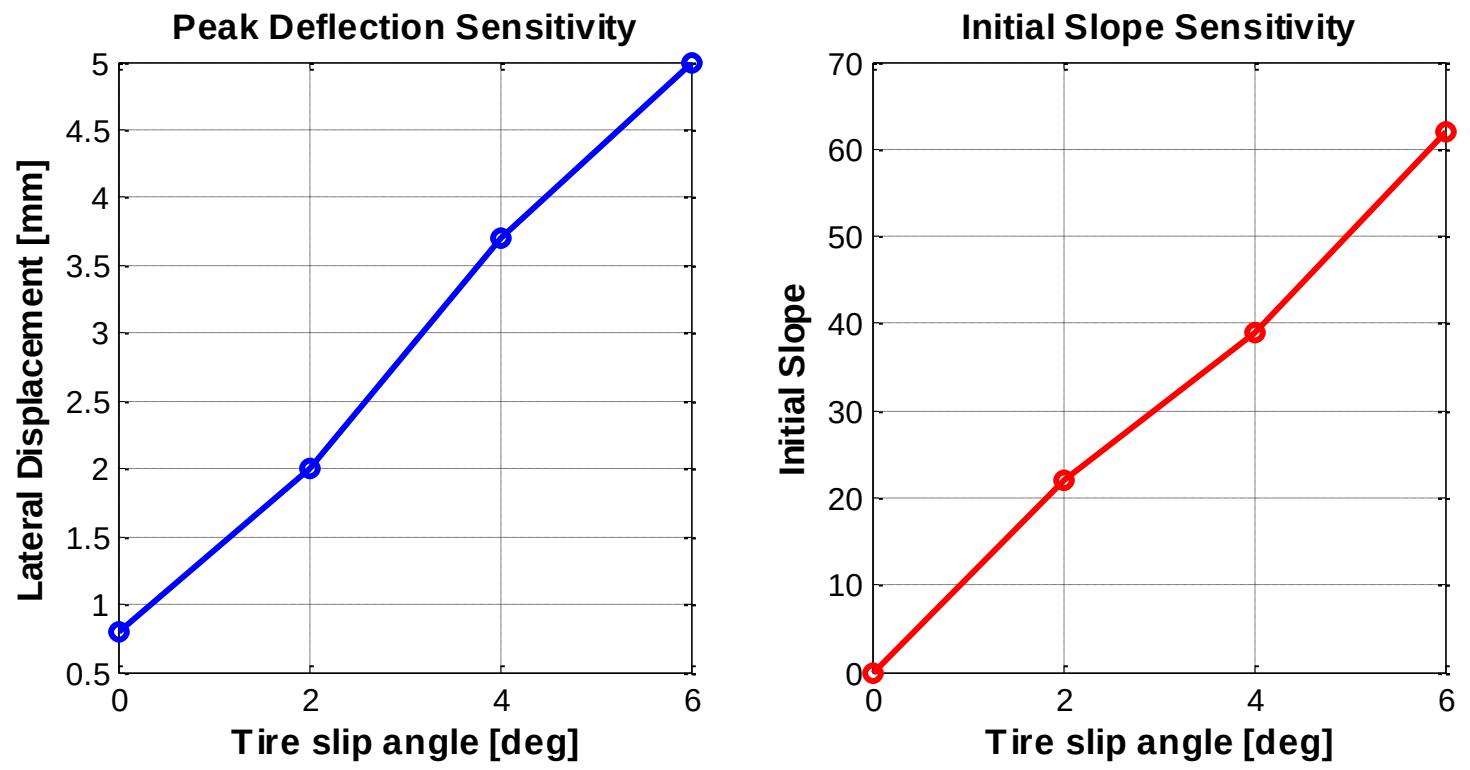

**Figure 13: Correlation between extracted features and the tire slip angle**

A multiple linear regression model was trained using both these signal features and found to estimate the tire slip angle with an accuracy of +/- 0.2 deg.

## 5. Conclusion

This paper provides a review of relevant works about the different signal processing techniques proposed by researchers to extract tire load and slip angle information from tire mounted accelerometers. Details about research activities undertaken as part of this study to develop a smart tire are presented. Novel algorithms for tire load and slip angle estimation are presented. Tire load is estimated by extracting the peak radial deflection of the tire from the radial acceleration signal. Tire slip angle is estimated by extracting the peak lateral deflection and the slope of the lateral deflection curve from the lateral acceleration signal. Future work will focus on using tire load and slip angle information to enhance the robustness of the vehicle state estimators.

Key challenges for the large-scale industrialization of Smart Tire Technology are as follows:

- For the installation of sensors in a tire, many problems will need to be considered, such as compatibility of the sensors with tire rubber i.e. stiffness issues, robustness of sensor in the harsh tire environment, wireless transmission of gathered data in an energy efficient manner, economic issues relating to the use of expensive sensors in a comparatively inexpensive product, the tire and finally meeting the power requirements of all the electronic components.
- Most studies have assessed the benefits of a smart tire for control systems in a simulation environment. It remains to be foreseen how of much of this benefit can be captured in the real-world, considering that most smart tires use a single point sensing solution transmitting data at a low frequency.
- It is also concluded that the current tire pressure monitoring chips available in production vehicles cannot support complex signal processing algorithms required to extract signal features from tire mounted sensors. Hence, an application specific IC would to be required.